\documentclass{emulateapj}

\newcommand{\kepler}{{\it Kepler}}

\newcommand{\prot}{\ensuremath{P_{\rm rot}}}
\usepackage{graphicx}
\usepackage{amsmath}
\usepackage{color}
\usepackage{natbib}
\usepackage{comment}

\shorttitle{Stellar Rotation and Orbital Periods of the Kepler Objects of Interest}
\shortauthors{McQuillan, Mazeh \& Aigrain}

\begin{document}

\title{Stellar Rotation Periods of the Kepler Objects of Interest: \\
A Dearth of Close-in Planets around Fast Rotators}

\author{A. McQuillan, T. Mazeh} 
\affil{School of Physics and Astronomy, Raymond and Beverly Sackler, Faculty of Exact Sciences, Tel Aviv University, 69978, Tel Aviv, Israel}
\email{amy@wise.tau.ac.il}
\author{S. Aigrain}
\affil{Department of Physics, University of Oxford, Oxford, OX1 3RH, UK\\}

\begin{abstract}
We present a large sample of stellar rotation periods for \kepler\ Objects of Interest (KOIs), based on three years of public \kepler\ data. 
These were measured by detecting periodic photometric modulation caused by star spots, using an algorithm based on the autocorrelation function (ACF) of the light curve, developed recently by McQuillan, Aigrain \& Mazeh (2013). Of the 1919 main-sequence exoplanet hosts analyzed, robust rotation periods were detected for 737. Comparing the detected stellar periods to the orbital periods of the innermost planet in each system reveals a notable lack of close-in planets around rapid rotators. It appears that only slowly spinning stars, with rotation periods longer than 5--10 days, host planets on orbits shorter than 3 days, although the mechanism(s) that lead(s) to this is not clear.
\end{abstract}


\section{Introduction}

A star-planet system has three different periods: the planetary orbital period and the stellar and planetary rotation periods. While the orbital period of the planet is almost the first thing we learn about the system when an exoplanet is discovered, the other two periods are not so easy to observe. It is of great interest to derive at least the stellar rotation periods for some of the known exoplanets, and compare them statistically with the orbital periods. This comparison may provide an insight into the formation and evolution of those systems.

Data from the \kepler\ mission are revolutionizing the study of both exoplanets and their host stars, with over 3000 exoplanet candidates identified to date \citep{bat+13}. \kepler's almost uninterrupted sub-mmag precision time series photometry allows for the derivation of precise orbital parameters from transit fitting, and stellar rotation measurements from star spot modulation \citep[e.g.][]{bas+11, mei+11,  nie+13}. \cite{sch+10, hir+12} have used the rotation periods of exoplanet host stars to study spin-orbit misalignment, through measurement of $v \sin i$, to study formation and evolution.

These studies all use Fourier-based methods for rotation period detection, while we use an autocorrelation function technique (ACF,  \citealt{mcq+13}), which we consider to be more robust. We applied the ACF to 1942 main-sequence KOIs and detected rotation periods in 737 targets that were found to host exoplanets. We then compared the rotation period to the orbital period of the inner-most planet of each system, showing a paucity of close-in planets around fast rotating stars. \\

\section{Data}
\label{sec:data}

We used the list of the KOIs and their parameters from the NASA Exoplanet Archive\footnote{http://exoplanetarchive.ipac.caltech.edu} (NEA, \citealt{ake+13}) of 22nd May 2013. Targets identified as false positives by either the \kepler\ pipeline or the NEA were excluded, providing a final list of 2010 stars.  These are listed in Table~\ref{tab:pers}, which appears in full in the online supplement. The effective temperatures and surface gravities used in this work come for the \kepler\ Input Catalog (KIC) \citep{bro+11} or where available from \cite{dre+13}. For 47 stars in the sample that are missing KIC values for $\log g$ and $T_{\rm eff}$, we used those provided in the NEA. These objects have the flag `N' in Table~\ref{tab:pers}.

Based on the $\log g$ and $T_{\rm eff}$ cut of \cite{cia+11}, 68 stars were identified as likely giants (flagged as `G' in Table~\ref{tab:pers}), and not included in this study. This leaves 1942 main-sequence targets.

The public release 14 quarter 3--14 (Q3--Q14) light curves for these targets are publicly available and were downloaded from the \kepler\ mission archive\footnote{http://archive.stsci.edu/kepler}. Q0 and Q1 were omitted due to their short duration, and Q2 was omitted due to significant residual systematics. We used the data corrected for instrumental systematics using PDC-MAP \citep{smi+12, stu+12}, which removes the majority of instrumental glitches and systematic trends using a Bayesian approach, while retaining most real (astrophysical) variability.

\section{Rotation Period Measurement}
\label{sec:rot_meas}

The rotation period measurement was performed using the autocorrelation function (ACF) technique, described in \cite{mcq+13}. This method measures the degree of self-similarity of the light curve over a range of lags. In the case of rotational modulation, the repeated spot-crossing signatures lead to ACF peaks at lags corresponding to the rotation period and integer multiples of it. We adopt this method of period detection over Fourier-based methods since the ACF has been shown to produce clear and robust results even when the amplitude and phase of the photometric modulation evolve significantly, and when systematic effects and long term trends are present.

\begin{figure}
  \centering
  \includegraphics[width=\linewidth]{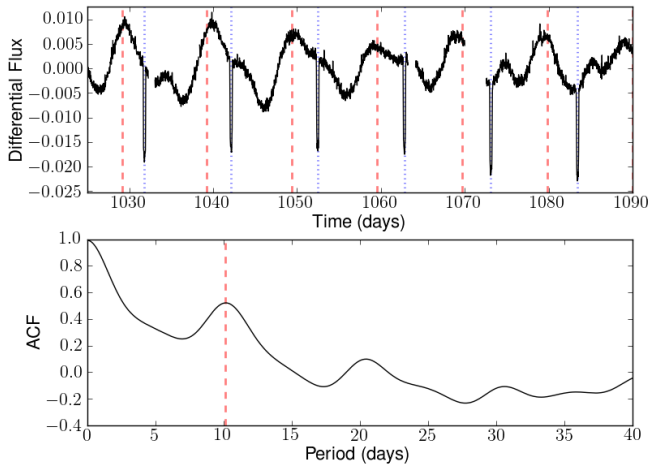}
  \caption{An example section of median normalized light curve from KOI\,805 (KIC\,3734868) (top panel), and the ACF of the full Q3-14 light curve (bottom panel). The orbital period (10.32 days) is marked with blue dotted lines on the light curve, where the transits are clearly visible. The detected rotation period (10.14 days) is marked on the ACF with a vertical dashed line and the corresponding period intervals are marked on the light curve as red vertical dashed lines. It can be clearly seen that the rotation period detected matches the repeating flux modulations.}
  \label{fig:lcex}
\end{figure}

We preprocessed the light curves following the method of \cite{mcq+13}, which includes median normalization of each quarter and masking the transit events. Minor modifications to the original code mean we now define the period as the gradient of a straight line fit to the ACF peak positions, as a function of peak number. Up to 4 consecutive peaks at near integer multiples of the selected ACF peak are used. This provides a more accurate period measure and uncertainty than the median and scatter of the peaks, which were previously used. 

We visually examined the ACF results, and any period detected was considered robust if repeated features are visible in the light curve at that period. For a more detailed discussion of this process, see \cite{mcq+13}. An example light curve and ACF are shown in Figure~\ref{fig:lcex}. 

Our analysis yielded period detections in 760 targets. A literature search and additional tests allowed us to exclude a further 23 targets from our sample: 17 targets were listed as confirmed or likely eclipsing binaries (EB, flag: `E') \citep{san+12, cab+12, mur+12, maz13}.  We found 4 that showed signs of binarity (secondary eclipses, eclipse depth differences, reflection/ellipsoidal effects (flag: `T'). One is a blended EB (flag: `B') and the centroid motion of one target on the \kepler\ CCD showed that the transits and rotation were on different targets (flag: `C'). We are left with 737 period detections for the remaining 1919 main-sequence planet hosts, upon which this study focuses.

The detected periods are between 0.9 and 62 days, and the amplitude of the modulation, $R_{\rm var}$, ranges from $\sim\,0.18$ to $\sim\,64$ mmags, where the \emph{range} $R_{\rm var}$ is defined as the interval between the 5th and the 95th percentiles of normalized flux per period bin.

\begin{figure*}
  \centering
  \includegraphics[width=0.8\linewidth]{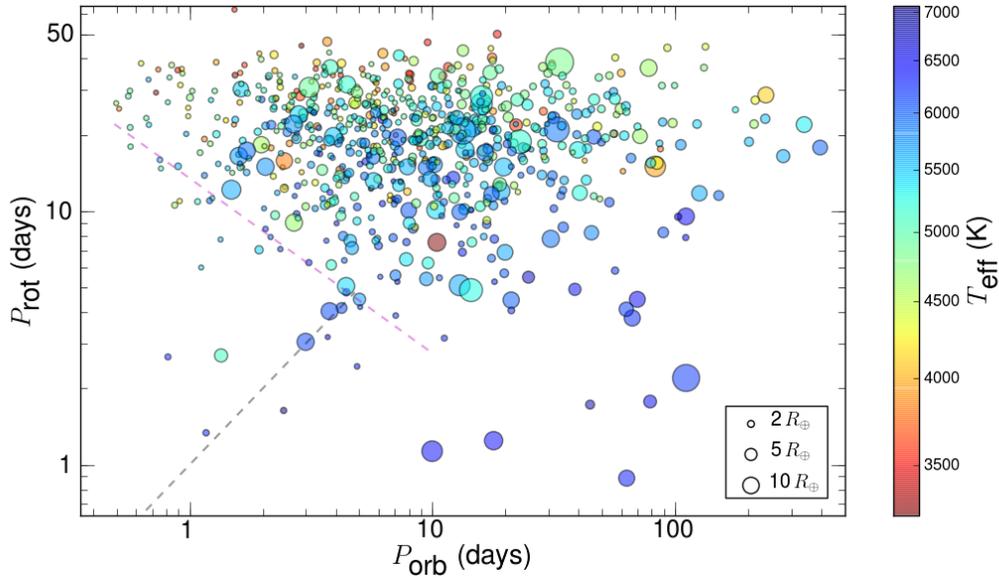}
  \caption{Stellar rotation as a function of orbital period of the innermost planet for the 737 KOIs with measured rotation periods. Point size scales as planet radius squared and color scales as $\log T_{\rm eff}$. The magenta dashed line is a fit of the lower envelope of points (see text). The black dashed line marks 1:1 synchronization between $P_{\rm orb}$ and $P_{\rm rot}$.}
  \label{fig:prot_porb_main}
\end{figure*}

\section{Results}
\label{sec:results}

The comparison of stellar rotation period, $P_{\rm rot}$,  to the orbital period, $P_{\rm orb}$, of the closest observed planet is shown in Figure~\ref{fig:prot_porb_main}, with planet radius indicated by the point sizes, and stellar effective temperature shown by color scale. The notable feature that emerges is the dearth of planets at short orbital periods around fast rotating stars; only slow stellar rotators, with periods longer than 5--10 days, have planets with periods shorter than 3 days. 

To assess the significance of this difference we performed a two sample Anderson-Darling test \citep{sch+87} on the orbital period distributions for $P_{\rm orb} \leq 10$\, days, divided at $P_{\rm rot} = 20.15$\, days. This division provides an equal number of systems (216) in each set. The p-value for the null hypothesis that both samples come from the same distribution is 0.018.

The dashed magenta line shows a fit to the lower edge of the distribution. This fit was obtained using points in the region bounded by $P_{\rm orb} \leq 10$ and $P_{\rm rot} \geq 3$. The lower bound on $P_{\rm rot}$ removes 6 outliers which do not form part of the main distribution of points. To find the lower envelope we minimized a function that incorporates the perpendicular distance between the line and points, and uses different weights for points above and beneath the line:
\begin{equation}
\label{eq:min_fit}
\kappa(c,m) = \sum_{i=1}^{N}{\sigma_i} \left[ \frac{|y_i - (c+mx_i)|}{\sqrt{1+m^2}} \right]\,,
\end{equation}
where $c$ and $m$ are parameters of the line $l = c + mx$ to be found, and $\sigma$ is defined by
\begin{equation}
\label{eq:sigma}
\sigma_i = \begin{cases} \sigma_{-}, 
& \mbox{  if } y_i > c+mx_i  \\ 
\sigma_{+}, & \mbox{  if } y_i \leq c+mx_i  \end{cases}\, .
\end{equation}
In this case, values for $\sigma_{+}$ and $\sigma_{-}$ were selected to be 0.11 and 0.01 respectively, such that 95\% of the points are above the line.. Best fit values of $c = 1.13 \pm 0.02$ and $m = -0.69 \pm 0.07$ were found. We obtained uncertainties by performing the fit using a random selection of 80\% of the points, over 1000 iterations. 

The temperature scale shows that, in agreement with field star rotation studies \citep[e.g.][]{bou+13}, the hotter stars display typically shorter rotation periods. 

\begin{figure}
  \centering
  \includegraphics[width=\linewidth]{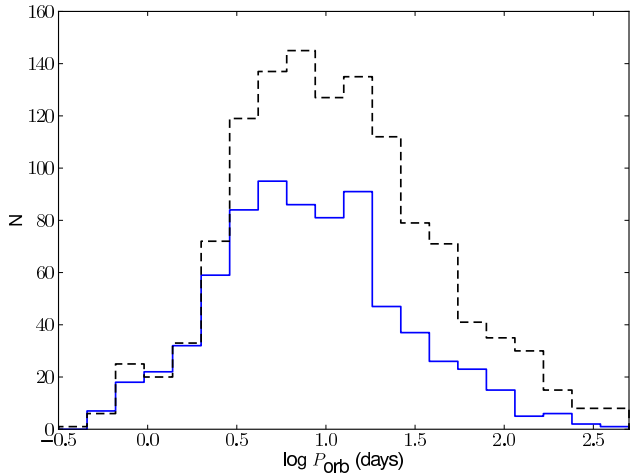}
  \caption{Distribution of orbital periods for the 737 KOIs with detected rotation periods (solid line) and 1182 without (dashed line), from the 1919 main-sequence exoplanet host stars.}
  \label{fig:per_nonper}
\end{figure}

Since Figure~\ref{fig:prot_porb_main} shows only the KOIs where a rotation period was detected, we compared the distribution of their orbital periods to those around stars without rotation measurements (see Figure~\ref{fig:per_nonper}). It is interesting to note that for the shortest orbital periods ($P_{\rm orb} < 2.5$ days), the number of period detections and non-detections is approximately equal, whereas at longer orbital periods there are far more non-detections.\\\\\\

\begin{table*}
  \caption{Stellar rotation periods for the KOIs.}
  \label{tab:pers}
  \centering
  \begin{tabular}{cccccccccc}
\hline
    KOI & KIC & $T_{\rm eff}$ & $\log g$ & $R_{\rm pl}$ & $P_{\rm orb}$ &  $\prot$\tablenotemark{a} & 
    $\sigma_{\rm  p}$\tablenotemark{a}  & $R_{\rm var}$\tablenotemark{a} & Flag\tablenotemark{b} \\
      &   & (K) & (g/cm$^3$) & ($R_{\oplus}$) & (days) & (days)  & (days) & (mmag) &\\ 
      \hline\hline
      	3&10748390&4766&4.59&4.68&4.888&29.472&0.134&11.75&N\\
	12&5812701&6419&4.26&13.40&17.855&1.245&0.124&0.78&-\\
	41&6521045&5909&4.28&1.24&6.887&24.988&2.192&0.39&-\\
	42&8866102&6170&4.10&2.71&17.834&20.850&0.007&1.12&-\\
	44&8845026&6250&3.50&9.61&66.468&3.792&0.907&1.16&-\\
        \hline\\
  \end{tabular}
  \tablecomments{
  This table is available in its entirety, in a machine-readable form
  in the online supplementary material. A portion is shown here for guidance 
  regarding its form and content.\\ 
  $^a$ Non-detections have `nan' in the $\prot$, $\sigma_{\rm  p}$ and $R_{\rm var}$ columns.\\
  $^b$ Flag definitions: N - KIC values missing so NEA values used; G - likely giant;  E - published EB; B - published blend;
  T - likely eclipsing binary identified in this work; C - centroid motion shows transit and rotation modulation on different stars.\\}
\end{table*}

\subsection{Caveats and Checks Performed}
\label{sec:caveats}

There are a number of caveats we considered in the analysis. There is a potential for contamination by periodic fluctuations in the light curve caused by phenomena other than rotation. Typical p-mode pulsations for main-sequence stars have periods on the order of minutes to a few hours \citep{nie+13}, which is below the shortest period detected in the KOI sample. The visual examination stage of the period confirmation also allows us to verify that the modulations appear typical of star spots.

Another potential cause of erroneous period detections are orbital effects (reflection, ellipsoidal variation) resulting from a massive planet close to the host star. This could affect the objects located in the lower left part of Figure~\ref{fig:prot_porb_main}, on the 1:1 rotation-orbit line. Most of the excluded EBs were located on this line, but those remaining do not show conclusive signs of binarity. Visual examination of their light curves confirms the existence of a modulation distinct from any possible orbital effects (slightly different period, evolving phase and amplitude), which is most naturally explained as (almost) synchronous rotation.

We have selected only the closest detected planet in each system for orbital period comparison. We obviously cannot rule out the possibility that additional undetected planets may exist with shorter orbital periods than the closest observed planet.  

Faster rotators are more variable on transit timescales, compared to stars with longer rotation periods. For all except the hottest stars, as the rotation period decreases the level of photometric variability increases. Therefore, planets may be harder to detect around fast rotators due to the increased activity signal which, if not effectively removed, can be problematic for transit detection algorithms. However, the detection probability of close-in planets is higher than for those on longer orbits, due to the larger number of transits observed in a given observation timespan, and the larger fraction of points in the transit. Therefore, the transit detection bias affecting fast rotators would not explain the paucity of close-in planets around them, since planets are observed further out around stars of the same rotation period, where they should be harder to detect. 

\section{Discussion}
\label{sec:disc}

We assume that most star-planet systems start their life in the lower-right corner of the $P_{\rm orb}$--$\prot$ diagram, with long orbital periods \citep{for+13} and fast rotations \citep{bou+13}. In the framework of this assumption, the planetary orbit of each system shrinks through inward migration, moving to the left in Figure~\ref{fig:prot_porb_main} \citep[e.g.][]{ras+96, war97a, fab+07, lub+11, for+13}, and the stellar rotation slows down, probably via magnetic braking operated by stellar wind, moving upwards in Figure~\ref{fig:prot_porb_main} \citep{kaw88}. The efficiency of this rotational breaking mechanism is dependent on the depth of the convective envelope, such that the most massive stars with very shallow convective envelopes remain fast rotators throughout their main-sequence lifetimes. The mass range (from KIC $T_{\rm eff}$ using stellar evolution models of \citealt{bar+98}) covered in this periodic sample is 0.3 -- 1.5\,$M_{\odot}$, leading to a wide range of spin-down rates.

The details of the rotation and orbital evolution, which depend on the physical parameters of the star, planet, accretion disc and probably the multiplicity of the system, determine the starting point and the track of each system in the $P_{\rm orb}$--$\prot$ parameter space. The ending points of the different tracks are determined by the halting mechanism(s) of the planetary migration \citep{ter03, daw+13, pla+13}, and the final stellar rotational period. Finally, in the last phase of their evolution, tidal interaction for short enough orbital periods might shift some of the systems, probably {\it towards} 1:1 spin-orbit synchronization \citep{maz08, zah08}. Figure~\ref{fig:prot_porb_main} presents the end product of the sample we have at hand --- the \kepler\ KOIs with detected stellar rotation, reflecting the final angular-momentum status of these systems.  Therefore, the $P_{\rm orb}$--$\prot$ diagram encodes valuable information on the formation and evolutionary tracks of different systems.
 
Figure 2 indicates that only slow stellar rotators, with periods longer than 5--10 days, host planets with periods shorter than 3 days. This unexpected state of affairs could be a result of tidal interaction, but planetary migration and the stellar braking process could also play a role in preventing systems from ending up in this zone of avoidance.
 
One factor which could link the stellar rotation to the orbital period of the innermost planet is the strength of the star's magnetic field. It can determine the radius of the cavity of the accretion disc in the pre-main-sequence phase, on one hand, and the angular momentum loss rate on the main sequence, on the other hand. The former might determine the point where the planetary migration stops \citep{kle+12}, while the latter determines the stellar rotation evolution. However, in its simplest form, this argument would suggest that a larger magnetic field leads to longer orbital periods {\it and} longer stellar rotation period, which is not what the systems in the top-left corner of Figure~\ref{fig:prot_porb_main} show. The present very short orbital periods could have been determined by other parameters of the system, but they would still need to be associated with the stellar angular momentum to explain our results.
 
Detailed theoretical models are needed to understand the evolution of the angular momentum of star-planet systems \citep[e.g.][]{mor+09}. Figure~\ref{fig:prot_porb_main}, together with the emerging data on the relative alignments of the stellar rotational axis and orbital plane in many systems \citep[e.g.][]{alb+12, alb+13}, might constitute a valuable input for such models.

\acknowledgments
\noindent {\it Acknowledgments:} The authors wish to thank Shay Zucker for his help with the envelope edge detection algorithm, and Simchon Faigler and Tomer Holczer for their tests to identify EB contamination. The authors also thank Dan Fabrycky, Darin Ragozzine and Jack Lissauer for helpful comments on an earlier version of this manuscript. TM is indebted to Arieh Konigl for illuminating discussions on the role of stellar magnetic fields and accretion discs. The research leading to these results has received funding from the European Research Council under the EU's Seventh Framework Programme (FP7/(2007-2013)/ ERC Grant Agreement
No. 291352). SA wishes to acknowledge support from STFC Consolidated grant ref. ST/K00106X/1 and Leverhulme Research Project grant RPG-2012-661. This research has made use of the NASA Exoplanet Archive, which is operated by the California Institute of Technology, under contract with the National Aeronautics and Space Administration under the Exoplanet Exploration Program. All of the data presented in this paper were obtained from the Mikulski Archive for Space Telescopes (MAST). STScI is
operated by the Association of Universities for Research in Astronomy,
Inc., under NASA contract NAS5-26555. Support for MAST for non-HST
data is provided by the NASA Office of Space Science via grant
NNX09AF08G and by other grants and contracts.  

\bibliographystyle{apj}



\end{document}